\documentclass[12pt]{article}
\begin{document}
\subsection*{Physics in a diverse world}
\subsubsection*{or A Spherical Cow Model of Physics Talent}
\subsubsection*{Talk at the KITP Conference: Snowmass Theory Frontier}
\subsubsection*{Feb. 23, 2022}
\subsubsection*{Howard Georgi}

I suspect that many of you, when you think about diversity in physics
feel as I did for much of my life.  We say to ourselves something like
``I am a good person and I would like to
increase diversity in physics but we
face societal obstacles not of our own making
and there is nothing I can do.''

I don't want this talk to sound judgmental and I certainly don't want to hold
myself up as a role model. In fact,
I am very embarrassed that it took me almost 45 years to realize that 
assuming there is nothing I can do 
is both morally untenable and very bad for physics.  For me, the
epiphany came when I was chairman of the Harvard Physics Department in 1992
and it had to do with gender diversity.
By this time I had been teaching at Harvard for over 15 years and I was very
spoiled.  

I thought that my job as a physicist was just to do good physics.
Helen Quinn was at Harvard when I arrived as a postdoc and we had many
discussions about the dearth of women in physics.  I knew it was a problem
but I didn't see how I could help.   

I thought
that my job as a teacher was just to organize a subject in the deepest and
most interesting way (which I enjoyed), come up with devilishly clever
problems (which I also enjoyed) and give lectures (which I was never very good at).
I was lucky to be at an institution where I could get away with this and where
there were always some students who were excited by the challenge of 
assimilating my oracular pronouncements and struggling with my crazy problems.
Through no particular fault of my own, I had many
spectacular graduate students whom I thought of
as just great young colleagues.  Our high energy theory group in the '80s included
Shelly Glashow and Steve Weinberg, so we had our pick of students who
wanted to do particle phenomenology, and many of them ended up working for
me, including four amazing women: Sally Dawson, Ann Nelson, Lisa Randall
and Liz Simmons.  Ann Nelson's graduate class in particular was very small
with several brilliant women and other diverse and interesting characters.
This class had a big positive influence on the culture of the department.
For example, they contributed to the mental health of graduate students by
initiating the ``Puppet Show'' in which the second-year students made anonymous
fun of the faculty (each of us represented by a silly puppet) to give the first-year
students the real scoop about the department.  This wonderful tradition is
still going on.

From that time on, I spent a lot of time on the graduate admissions
committee trying to ensure that the women applicants were treated fairly.
The number of racial minorities in the applicant pool was still almost zero.
But when the women graduate students
tried to explain that there was a real problem for women
in physics, I still
did not get it.  

Meanwhile I was having a wonderful time teaching
undergraduate courses 
as well as my graduate courses.  Many undergrads hung out in
my office and I got to know many of them, men and women, very well.
There were some amazing kids.  Again, some
of the women tried to explain how difficult the physics culture was for
them, but it just wasn't getting through my thick skull.

What happened in 1992, to make a long story short, was that I had some data
broken down by gender that showed that the young men who graduated in physics
loved their time in the Harvard 
Physics department and the women hated it!  And these were
women who loved physics so much that they stuck it out and graduated in
physics in spite of the fact that they felt like outcasts.  Finally I
understood what the women had been telling us.  They were in an abusive
relationship with the Harvard Physics Department.  This was just not right!

Ever since that time, I have done my best to encourage diversity.  The
effort has been extremely rewarding.  I will say a bit more about it in a
few minutes.

So what is wrong with saying ``there is nothing I can do''?

We absolutely do face societal obstacles and we must do everything in our
power to break them down. Many children who could develop into outstanding
scientists cannot even imagine a
career as a physicist and far too many minority kids would not be able to
afford to pursue a physics career even if they dreamed of it.  And it is an
understatement to say that one of our major
political parties is not trying to change this.   I know that many of
you are working hard with your outreach and your teaching to chip away at
this problem.

But we also put up a daunting obstacle of our own making.  Perhaps
subconsciously, we are drawn to a metaphor of physics as 
survival of the fittest and we look for the apex predator who will
aggressively claw a way to the top of the food chain.  So in our teaching
and our admissions and in our hiring and most importantly 
in our own heads we make ordered
lists and we search for ``the best.''  

This is not surprising.
We are drawn to physics because we want to give real answers
to real questions. We don't make up our questions. We try to uncover how
the world works at the deepest levels. And we don't make up our answers.
We are not satisfied 
unless our understanding is quantitative and expressible in the language of
mathematics and does not depend on how cleverly
we express it in human language.  
And we submit our answers to the ultimate test - quantitative comparison
with experiment.
So naturally, we tend to
quantify our thinking about physicists. 
Then we say that if white male
applicant A is higher on our ordered list than minority
applicant B we must choose A
and there is nothing we can do about diversity.  This
is absolute nonsense!

I am absolutely ABSOLUTELY
NOT suggesting naive affirmative action.  Few things damage
the cause of diversity more than choosing a minority applicant who is not
outstanding.  
The first requirement for any candidate is that they must be able to do a
great job.  The trouble is that for physics, one doesn't
know exactly what the job is and
the assumption that we can unambiguously order
outstanding people makes no sense.  We physicists should know
better than anyone that there are quantitative questions that do not have
answers, like which of two space-like separated events comes first.

In fact, I believe, there is more to this issue than just removing
obstacles to diversity.  The right metaphor, I believe, gives us a powerful
incentive to increase diversity in physics --- not just because it is the
right thing to do, but because it is important for physics.
Here I want to talk to you established physicists out there. I know that each
of you
has incredible skills that you have worked hard to hone and use
productively.  But I also believe strongly that each of you has a DIFFERENT
cocktail of skills.  If you are good at people-watching, you will know this.
I see this myself in many ways.  Looking inward, I know that there 
are some things that I am MUCH better at than other things. Looking
outward, I see a similar incredible 
diversity of skills in my students and my younger colleagues. And looking
back on my long career, I have been fortunate to work with many 
great physicists in my own subfield of particle theory.  
I have had the privilege of collaborating closely with almost ninety amazing
physicists, including dozens of winners of various physics prizes.
And of course I have gotten to know
many people very well without actually collaborating.  I shared an office
for a number of years 
with Ed Witten which was a humbling experience. All of these
people are simply amazing intellects, but what I have observed is that
each one is amazing in a 
very unique way.  Here I do not mean how they talk about physics.  All
of these characters speak the language of physics, whatever natural
language they are comfortable with.  But if you get to know them well
enough to glimpse what is going on in their heads, you see wild diversity.

The conclusion I draw from many years of physicist
watching is that if you want to quantify what makes a great physicist, you
must use a space with very many dimensions, a different dimension for each
of the very many possible ways of thinking that may be important for really
interesting problems.  I sometimes imagine a
spherical cow model of physics talent in N dimensions where N is
large and talent in each dimension increases from 0 at the origin to 1 at
the boundary, the N-dimensional version of the positive octant of a sphere.
This, I learned from Wikipedia, is called an ``orthant''. 
Each point in my N-dimensional orthant is a possible set of
talents for physics. Great physicists are out near the boundary, far away
from the origin. If we
assume that the talents are uniformly distributed, you can see that in my
spherical cow model, the fraction of possible physics talents within $\epsilon$ of
the boundary grows like
N times $\epsilon$ for small $\epsilon$.  If N is very large, as I think it is, that means
that there is a lot of space near the boundary!  
What this suggests to me is that there are
a huge number of ways of being a great physicist and that 
in turn suggests there are many ways
of being a great physicist that we haven't seen yet.  Diversity is critical
to the future of physics, I believe, because it is imperative to explore
this vast space of physics talent and that means encouraging people who are
different and who think and act differently.

You can quibble with the details of my spherical cow orthant, but I
will be very happy if this helps some of you recognize how damaging it is
to rely on one dimensional ordered lists in dealing with people.  If such a
list means anything at 
all, it represents only some arbitrary one-dimensional projection from some
much higher dimensional space.  
This is why I think that working for diversity in physics is not just a moral
imperative, it is the most useful approach for our groups, for our
institutions, and for the field of physics that we all love.  To do this
effectively, you should keep the spherical cow in mind.

In your teaching, you probably have to give grades, but I think you should give
your students sub-grades for several ways of excelling in your courses and
you should keep track of 
each of them separately.  This makes it much easier to
encourage and get to know a diverse 
set of students and to explain that the final grade is a somewhat arbitrary
combination.  You should get to know your students as people and
celebrate their uniqueness and try to encourage them to find and recognize
and be proud of their particular strengths.

In your mentoring, I believe that you should not try to force a younger colleague into a
predetermined mold but instead encourage them to develop their unique
strengths and help them to understand how to display them most
effectively.  Sometimes this also means working to help your older
colleagues accept new ways of thinking.

The most important thing you can do for diversity is to hire diverse
faculty.  I have seen this effect dramatically for women in my own
department.
We still have a long way to go but having a group of outstanding
women faculty who are very different among themselves has made a big
change in the culture of the department and in the morale of the
students.  I could not have been happier when three of our women undergrads
won Rhodes Scholarships this year.  All of them are very active in our women in
physics group.  I look forward to the day when our department is equally
diverse racially and culturally.  I think it will happen if we 
work harder to
consider candidates as people with multiple skills rather than
numbers in a ranked list.

There are certainly some very rare super-super-stars who are so far out in some
direction in the space of physics talent that they are clearly unique.
If you can find another Ann Nelson or Ed Witten you should hire her independent of
her minority status. But most of our candidates, like most of us, will just
be ``ordinary'' good physicists who have a
combination of skills and, if they are lucky, those skills will sometimes be the
right skills to solve important problems.  
Selection committees need to
avoid overly rigid definitions of sub-field so they can search broadly.
They need to avoid confusing aggressiveness or facileness with ability.  
They need to apply an 
appropriate implicit bias factor for candidates who look exactly like the
selection committee and the current department.
They need to avoid relying on the ``old-boy network'' and to make a special
effort to identify promising candidates who have not come up through the
usual channels.
None of this is easy. 

At the national level, you need to 
educate yourselves about the outstanding minority physicists in your own
subfields.  Invite them to give important talks.
Nominate them for prizes.  

While I care deeply about these issues, I know that my own view is
narrow and personal and also probably outdated in many ways because the
landscape is changing with time.  So I am delighted that we have a panel of
three young physicists who will, I hope, give you a more current
perspective on these issues.  I am very grateful to them for being willing
to share their thoughts.

I can't give a talk like this without getting choked up at the thought of
the incalculable loss to Physics of my former student Ann Nelson.  Ann was
one of the most remarkable of the amazing people I have been privileged to
know.  I have said before that while I have had many students who are much
better than I am at many things, Ann was my only student who was 
much better than I am at what I do best.  But she was not only an amazingly
good physicist.   She was an amazingly good person.
I am very grateful to Devin for inviting me to honor Ann's legacy in this session.
Let me close by quoting from her backpage article in a 2017 Physics Today.

``If your career is established and you are not making an
explicit and continual effort to encourage, mentor, and support all young
physicists, to create a welcoming climate in your department, and to
promote the hiring of diverse faculty members, you are part of the
problem. This is a critical issue of civil rights in our field.'' 

I have tried to argue here that this is more than a civil rights issue.  It
is the best way to ensure that Physics will continue to be great.

So what is your assignment?

Your first job is to do great physics and enjoy it and communicate your
excitement to the next generation.  But you owe it to this field that we
love to work to increase its diversity. 

The most important thing is to keep at it!  This is a job for optimists.
Progress will always be slower than we would like.  That is just the way the
world works.  But progress will not happen at all unless the good people
who think that there is nothing they can do actually wake up --- and start doing.

\end{document}